# Surrogate Empowered Sim2Real Transfer of Deep Reinforcement Learning for ORC Superheat Control

Runze Lin [1,2], Yangyang Luo [1], Xialai Wu [3], Junghui Chen [4,*] ,
Biao Huang [2,*], Lei Xie [1], Hongye Su [1]

[1] State Key Laboratory of Industrial Control Technology, Institute of Cyber-Systems and Control, Zhejiang University, Hangzhou 310027, China

[2] Department of Chemical and Materials Engineering, University of Alberta, Edmonton, AB T6G 2G6, Canada

[3] Huzhou Key Laboratory of Intelligent Sensing and Optimal Control for Industrial Systems, School of Engineering, Huzhou University, Huzhou 313000, China

[4] Department of Chemical Engineering, Chung-Yuan Christian University, Taoyuan 32023, Taiwan, R.O.C.

**Preprint Submitted to *Elsevier***

**July 31, 2023**

* **Corresponding authors.**
**E-mail addresses:** jason@wavenet.cycu.edu.tw (Junghui Chen), biao.huang@ualberta.ca (Biao Huang)

**Abstract**

The Organic Rankine Cycle (ORC) is widely used in industrial waste heat recovery due to its simple structure and easy maintenance. However, in the context of smart manufacturing in the process industry, traditional model-based optimization control methods are unable to adapt to the varying operating conditions of the ORC system or sudden changes in operating modes. Deep reinforcement learning (DRL) has significant advantages in situations with uncertainty as it directly achieves control objectives by interacting with the environment without requiring an explicit model of the controlled plant. Nevertheless, direct application of DRL to physical ORC systems presents unacceptable safety risks, and its generalization performance under model-plant mismatch is insufficient to support ORC control requirements. Therefore, this paper proposes a Sim2Real transfer learning-based DRL control method for ORC superheat control, which aims to provide a new simple, feasible, and user-friendly solution for energy system optimization control. Experimental results show that the proposed method greatly improves the training speed of DRL in ORC control problems and solves the generalization performance issue of the agent under multiple operating conditions through Sim2Real transfer.

## 1. Introduction

In the context of Industry 4.0 and smart manufacturing, the transformation to intelligent processes is also necessary in the field of process control. The primary objective of industrial control is to effectively achieve energy savings, emission reduction, and efficiency improvement in industrial operations. The Organic Rankine Cycle (ORC) [1-3] is a promising and highly efficient technology for recovering and generating electricity from low-grade waste heat. By using various organic substances as working fluids, it can recover low-temperature waste heat in a wide range of environments. Due to its simple structure and high heat recovery efficiency, the ORC system has recently attracted significant attention in the field of industrial waste heat recovery.

However, the operating conditions of the waste heat typically have drastic changes during power generation, such as temperatures and flow rates. These fluctuations not only affect the operating conditions of the ORC system but also impact its dynamics, leading to significant variations in the waste heat. As a result, the ORC system becomes a complex thermodynamic process that involves the phase change flow of the working fluid [4, 5], resulting in highly nonlinear and time-varying dynamics. Although there have been numerous studies on ORC systems, most of them have primarily focused on aspects such as system design and parameter optimization. Unfortunately, there has been relatively limited research on ORC control strategies.

A good control system design can extend the life of the ORC system and improve the recovery and utilization of waste heat. Due to the fluctuating nature of the waste heat operating conditions, a proper control system is required to ensure the safe operation of the ORC system. An important performance indicator for the safety of the system is the superheat (SH) of the working fluid at the outlet of the evaporator [1]. To achieve the goal of maintaining the desired SH setpoint in this paper, the problem of the large variation in the expander's rotational speed must be addressed, as the changes in the heat source during operation can pose a great challenge to the controller design. By implementing a robust control system, the SH should be controlled effectively even in the presence of external uncertainties.

The control of ORC systems for waste heat recovery faces numerous difficulties and challenges. The thermodynamic properties of ORC systems are highly coupled and multivariable, and they exhibit time-varying nonlinearity while operating under fluctuating waste heat conditions. Many researchers have studied the control of ORC waste heat recovery systems. Common control schemes for ORC systems include Proportional-Integral-Derivative (PID) control, gain-scheduling PID control, Model Predictive Control (MPC), and their respective variations. Gain scheduling is the most common PID algorithm used in the industry. It adjusts the controller gains in local operating regions to overcome nonlinear process characteristics. This approach avoids the degradation of closed-loop performance to the extent of process instability that can result from the use of fixed-gain controllers [6]. Grelet et al. [2, 7] developed a gain-scheduling control strategy for on-board ORC waste heat recovery systems. Zhang et al. [3] applied gain-scheduling Linear Parameter Varying (LPV) robust controllers to ORC-based waste heat energy conversion systems. Peralez et al. [8]

studied an ORC system for waste heat recovery from heavy-duty diesel engines and proposed a control design that enables a two-level closed-loop control strategy, with an energy management system providing setpoints for the lower-level controller. Wu et al. [5] systematically investigated the integration of control and optimization in the ORC process, using Structured Singular Value (SSV) analysis for ORC operating parameter optimization and control integration. Additionally, model-based optimization control methods such as MPC have been extensively studied for the control of waste heat recovery ORC systems. Zhang et al. [9] developed a control-oriented model for an ORC-based waste heat energy conversion system and proposed a constrained MPC strategy. Hernandez et al. [10] used sparse identification techniques to derive a polynomial model used in the nonlinear MPC (NMPC) framework with nonlinear extended prediction self-adaptive control. Shi et al. [11] presented a dual-mode fast dynamic matrix control algorithm for ORC control. Wu et al. [12] addressed the wide-range fluctuation of waste heat conditions by using economic MPC (EMPC) to optimize the ORC system. They used a simplified ORC mechanism model as the prediction model in EMPC, effectively improving the waste heat recovery efficiency (net power output) of the ORC system. Zhang et al. [13] proposed an Adaptive Dynamic Programming (ADP) approach to control the ORC system. They addressed safety-related constraints based on basis functions and introduced an event-triggered mechanism to reduce computational costs during online controller operation.

Meanwhile, Deep Reinforcement Learning (DRL) is a very popular technique in both academia and industry, with extensive research being conducted recently. Attention towards DRL skyrocketed after AlphaGo's victory over human players in the game of Go [14]. Generally speaking, DRL addresses the problem of sequential decision-making in uncertain environments [15], which is well suited to the goals of process control, especially when significant uncertainties are involved [16]. Mathematically, DRL is formulated as an optimal sequential decision-making algorithm that can account for randomness within a system. Additionally, DRL not only "stores" memories of offline-trained control policies but also has the potential for adaptability and transfer learning [17, 18].

Unlike MPC, DRL-based control design is a kind of data-driven approach that learns from experience through trial and error and can handle complex systems with unknown dynamics by interacting with the environment; while MPC ~~still~~ requires a system model to predict future behavior and uses online optimization to calculate optimal control inputs. The data-driven model-free characteristics of DRL make it well-suited for problems where the system dynamics are difficult to model or where the system is subject to changes over time. Besides, it is also better suited for control problems with time-varying uncertainty.

In recent years, a limited number of studies have started to focus on the integration and application of DRL in the optimization and control of ORC systems. Wang et al. [19] investigated the use of DRL to control the SH in an ORC system, particularly during transient changes in heat sources. Their study demonstrated the successful implementation of a DRL-based PID controller and showed its reliable performance even in the presence of disturbances. A switching mechanism was also introduced to extend the operating range and improve the learning performance of the two controllers. In another study, Xu and Li [20] conducted a comparative performance analysis of various methods, including *Q*-learning, offline and online dynamic programming (DP), for online transient power optimization in ORC systems. The simulation results indicated that the *Q*-learning approach exhibited excellent optimization performance, making it a promising option for optimizing the system's power generation under transient conditions.

Next, the aforementioned DRL-based ORC superheat control method was discussed in detail [19]. Two approaches were used to control the SH in the ORC system. One approach was to use a DRL agent directly to control the SH, with the agent's action serving as the control signal to track the setpoint of the SH and eliminate errors. The other approach was to use a DRL agent to dynamically calculate and adjust the PID controller's parameters online, to address varying operating conditions and suppress disturbances caused by changes in waste heat. However, directly performing DRL training on the physical system is impractical since it often takes several days for a single training session. Although building an accurate mechanistic model for DRL training would be an alternative way, it is still pretty labor-intensive and computationally expensive. To solve the above problems, a new training method needs to be proposed.

There are still significant challenges to the practical application of DRL in ORC control, which is necessary to create value for businesses and society. First, the implementation of DRL remains largely confined to the realm of games, where infinite trial and error is possible. Direct interaction and training of DRL with actual ORC systems raises significant safety concerns and incur prohibitively high costs for trial and error, making direct training of DRL on physical ORC systems in industrial production impractical. Second, although the use of an accurate model to facilitate the interaction with DRL agents can mitigate safety issues, developing complex mechanistic models requires considerable effort, and training these models is computationally time-consuming with low sampling efficiency. Additionally, DRL exhibits poor generalization performance and is sensitive to model-plant mismatch, often resulting in DRL agents that can only make correct decisions within the specific environment in which they were trained, leading to significant performance degradation when slight environmental changes occur.

In summary, the above studies have not well considered the feasibility of applying DRL to process control, particularly overlooking the important factors of safety and generalization, which hinder the large-scale application of DRL in industrial scenarios. To tackle these challenges, this paper proposes a DRL-based ORC system controller training method that combines industrial big data modeling, transfer learning, and DRL. The goal is to address the safety issues of DRL in ORC system control, thereby improving training efficiency and control performance, and reducing training costs. Our alternative goal is to leverage the rich information embedded in historical operational data to further enhance the performance of ORC systems, thus providing new insights for the practical application of DRL in industrial process control and energy system engineering.

To the best of our knowledge, there is currently no existing research on surrogate-based and DRL-based controller training with simulation-to-real (Sim2Real) transfer in the field of ORC control. This paper is the first to address the ORC superheat control problem using the concept of DRL Sim2Real transfer learning, innovatively proposing the use of virtual prototypes for offline pre-training and subsequent Sim2Real transfer to the real system. The rest of this paper is organized as follows. Section 2 describes the ORC system and provides a systematic analysis of modeling and control. In the methodology section, Section 3 introduces the DRL-based Sim2Real control design method. Subsequently, in Section 4, experimental discussions are presented to provide specific implementation details of the methodology and analyze the experimental results. Finally, Section 5 summarizes the research content and contributions of this paper.

## 2. System description, modeling, and control

### 2.1. ORC system description

The ORC system, a widely adopted approach to power generation from low-grade heat sources, consists of four primary components: the evaporator, the condenser, the expander, and the pump. As shown in Fig. 1, the ORC system operates through a cyclic process [3, 5, 8]. Initially, the organic working fluid is transformed into superheated steam by absorbing heat from the heat source in the evaporator. Subsequently, the superheated steam is passed to the expander to generate power, where thermal energy is converted to mechanical energy through the expansion process. The low-pressure steam exiting the expander is then condensed back into liquid in the condenser. Finally, the liquid fluid is pressurized by the pump and returned to the evaporator, starting a new cycle in the ORC-based waste heat recovery process.

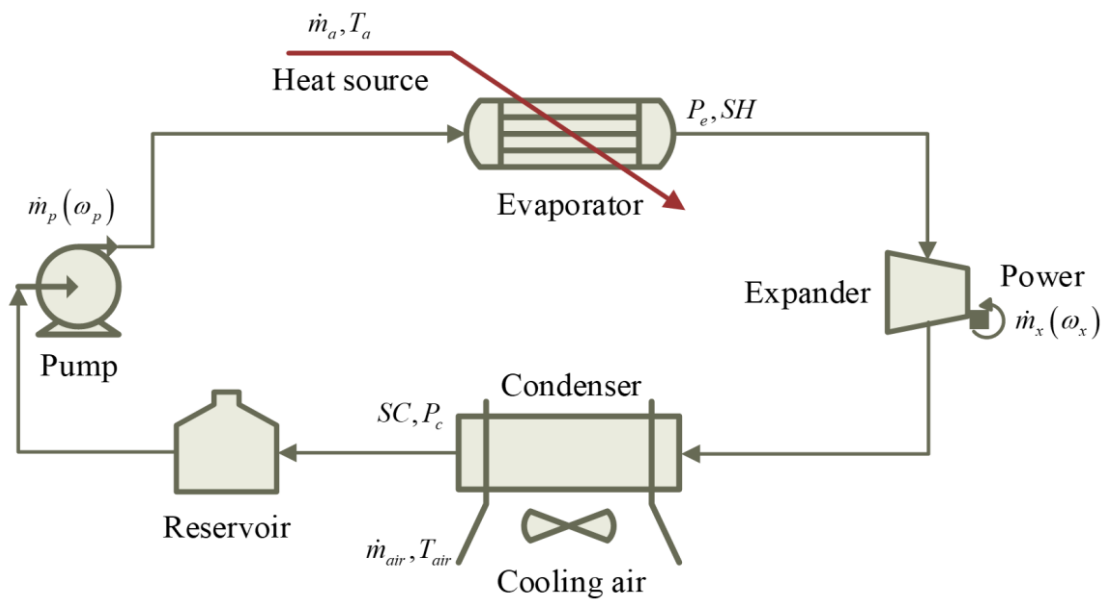

**Fig. 1 Schematic diagram of the ORC system.**

The integration of (1) phase change flow of the working fluid and (2) internal heat and mass transfer makes the ORC system a complex thermodynamic cycle with multivariate coupling and strong nonlinearity [5]. It is, therefore, necessary to dynamically model the ORC system according to its thermodynamic mechanism and characteristics. In this regard, a dynamic modeling method can be used to provide a systematic understanding of the ORC system behavior. To model the dynamic behavior of an ORC system, a set of differential equations is established based on the governing equations of the components and their interactions. These equations are solved numerically to predict the system dynamics and to develop optimal control strategies for enhanced performance.

## 2.2. Process dynamics modeling

### 2.2.1. Evaporator

The primary goal of an ORC system is to efficiently recover waste heat from industrial processes and convert it to electricity. The dynamic behavior of the entire system is significantly influenced by the evaporator [5, 12], making it the most critical component of the ORC setup. On the other hand, the pump and the expander can be adequately described by static algebraic equations. For this research, assume perfect condensation of the organic working fluid in the condenser, thus simplifying the model by neglecting the dynamics of the condenser. Specifically, the moving boundary (MB) method is used to develop the dynamic model of the evaporator. This approach involves a distributed lumped parameter model, which effectively balances computational complexity with high modeling accuracy [21, 22]. For a more comprehensive understanding of the ORC system model, readers can refer to our previous papers [5, 11, 17].

Fig. 2 shows a schematic diagram of the MB model for the evaporator. In this model, the subcooled organic fluid is introduced through the inlet of the evaporator, and then the superheated vapor is discharged from its outlet [5]. Industrial waste heat is used to heat the organic fluid and then enters the evaporator with random mass flow rates and initial temperatures. The evaporator can be divided into three distinct regions: the preheating region (sub-cooled region), the evaporation region (two-phase region), and the superheating region. For each of these regions, corresponding dynamic models are established to capture their behaviors. Here, $L_1$, $L_2$, and $L_3$ represent the lengths of the three regions, respectively.

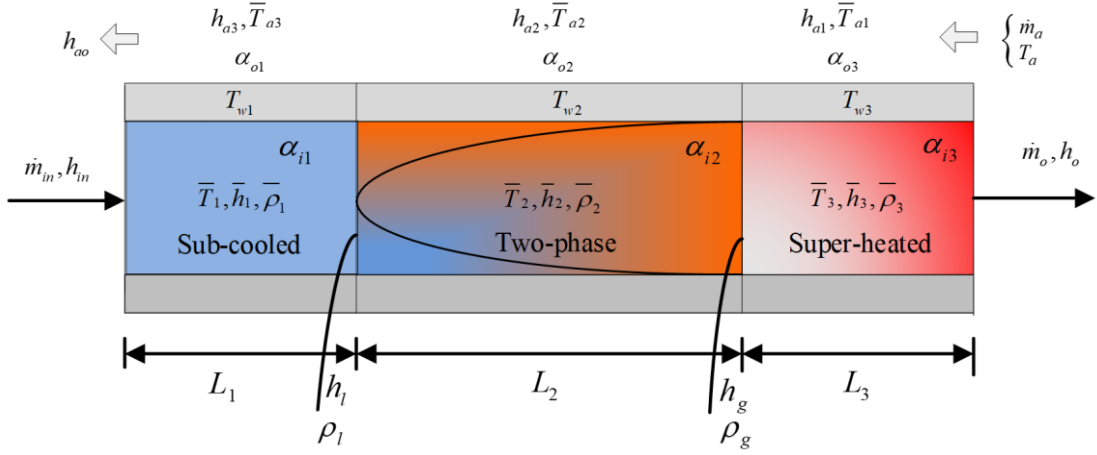

**Fig. 2 The MB model of the evaporator in the ORC system.**

Before deriving the model equations, certain simplifications and assumptions must be made for the evaporator model of the ORC system [23, 24]:

1. The evaporator is assumed to be a horizontally straight pipe, neglecting the influence of gravity terms.

2. The flow of the working fluid is assumed to be one-dimensional.

3. Axial heat conduction between the organic fluid and the tube is neglected, and heat transfer is assumed to occur only in the radial direction.

4. The pressure drop along the length is neglected, ignoring the conservation of momentum.

Based on these basic assumptions and the principles of conservation of energy and mass, the thermodynamic characteristics of the heat transfer process of the evaporator can be described by simplified first-order partial differential equations (PDEs). Overall, the evaporator model comprises two one-dimensional PDEs and one ordinary differential equation (ODE).

The mass balance of the working fluid for each region can be described by

$$\frac{\partial A \bar{\rho}_j}{\partial t} + \frac{\partial \bar{\dot{m}}_j}{\partial z} = 0, \, j = 1, 2, 3 \tag{1}$$

where $A$ represents the tube's cross-sectional area, $\bar{\rho}_j$ denotes the density of the working fluid, $\overline{\dot{m}}_j$ is the mass flow rate of the working fluid, and $_t$ and $z$ stand for time and axial length, respectively.

The energy balance of the working fluid for each region is governed by

$$\frac{\partial \left( \overline{\rho}_j A \overline{h}_j - AP \right)}{\partial t} + \frac{\partial \overline{\dot{m}}_j \overline{h}_j}{\partial z} = \pi D_i \alpha_{ij} \left( T_{wj} - \overline{T}_j \right), j = 1, 2, 3 \tag{2}$$

where $\overline{h}_j$ is the specific enthalpy of the working fluid, $D_i$ represents the tube's inner diameter, $\alpha_{ij}$ is the convective heat transfer coefficient on the inner surface of the tube, and $T_{wj}$ and $\overline{T}_j$ represent the temperatures of the wall and the working fluid, respectively.

The energy balance of the pipe wall for each region is described by

$$\left( C_p \rho A \right)_w \frac{\partial T_{wj}}{\partial t} = \pi D_i \alpha_{ij} \left( \overline{T}_j - T_{wj} \right) + \pi D_o \alpha_{oj} \left( T_{aj} - T_{wj} \right), j = 1, 2, 3 \tag{3}$$

where $C_p$ is the heat capacity, $D_o$ stands for the tube's outer diameter, $\alpha_{oj}$ is the convective heat transfer coefficient on the outer surface of the tube, and $T_{aj}$ is the temperature of the industrial waste heat at each section.

Considering the instantaneous fluctuation characteristics of the waste heat from industrial processes, which has a faster thermal response than the wall and working fluid temperatures, the average temperature of the waste heat in each region ($\overline{T}_{aj}$) is directly calculated using the static energy balance equation given by:

$$0 = L_j \pi D_o \alpha_{oj} \left( T_{wj} - \overline{T}_{aj} \right) + \dot{m}_a \left( h_{aj,in} - h_{aj,out} \right), j = 1, 2, 3 \tag{4}$$

where $\dot{m}_a$ is the mass flow rate of the waste heat, $h_{aj,in}$ and $h_{aj,out}$ denote the inlet and outlet heat enthalpies in the $j$-th region, respectively.

In general, the MB model assumes a uniform distribution of thermodynamic properties within each region and achieves dynamic modeling of the evaporator by varying the lengths of the three regions. Thus, the basic idea of the MB modeling approach is to integrate the mass conservation Eq. (1) and the energy conservation Eq. (2) along the axial length of the evaporator in the three regions and to simplify the equations using the Leibniz formula to obtain the MB models for each region. The expression of the Leibniz formula is given by

$$\int_{z_1}^{z_2} \frac{\partial f(z,t)}{\partial t} dz = \frac{d}{dt} \int_{z_1}^{z_2} f(z,t) dz - f(z_2,t) \frac{dz_2}{dt} + f(z_1,t) \frac{dz_1}{dt} \tag{5}$$

After the intermediate terms are rearranged and eliminated, a system of ODEs consisting of seven states can be obtained. The evaporator model of the ORC system can be formulated as a nonlinear state space model:

$$\mathbf{D}(\mathbf{x})\dot{\mathbf{x}} = \mathbf{f}(\mathbf{x}, \mathbf{u}_e) \tag{6}$$

where the system state vector is $\mathbf{x} = \left[ L_1, L_2, P_e, h_o, T_{w1}, T_{w2}, T_{w3} \right]^T$, and the manipulated variable vector is $\mathbf{u}_e = \left[ \dot{m}_{in}, h_{in}, \dot{m}_o, \dot{m}_a, T_a \right]^T$.

However, it is noted that the dynamic characteristics of the system variables in the commonly used seven-order model still exhibit distinct differences in time scales. Therefore, in this study, it is decided to transform the three slow dynamic differential equations in the original evaporator model into algebraic equations, thereby reducing the order of the evaporator model (Eq. (6)) from 7 to 4. Thus, the reduced-order evaporator model can be represented by a set of differential-algebraic equations (DAEs):

$$\begin{cases} \dot{\mathbf{x}}_r = \mathbf{f}_r(\mathbf{x}_r, \mathbf{y}_r, \mathbf{u}_e) \\ 0 = \mathbf{g}(\mathbf{x}_r, \mathbf{y}_r, \mathbf{u}_e) \end{cases} \tag{7}$$

where $\mathbf{x}_r = \left[ T_{w1}, T_{w2}, T_{w3}, L_1 \right]^T$ represents the vector of differential variables, $\mathbf{y}_r = \left[ L_2, P_e, h_o \right]^T$ is the vector of algebraic variables, and the nonlinear time-varying functions in Eq. (7) are defined as follows:

$$\mathbf{f}_r(\mathbf{x}_r, \mathbf{y}_r, \mathbf{u}_e) = \begin{bmatrix} \pi D_i L_1 \alpha_{i1}(\overline{T}_1 - \overline{T}_{w1}) + \pi D_o L_1 \alpha_{o1}(\overline{T}_{a1} - \overline{T}_{w1}) \\ \pi D_i L_2 \alpha_{i2}(\overline{T}_2 - \overline{T}_{w2}) + \pi D_o L_2 \alpha_{o2}(\overline{T}_{a2} - \overline{T}_{w2}) \\ \pi D_i L_3 \alpha_{i3} \overline{T}_3 - \overline{T}_{w3}) + \pi D_o L_3 \alpha_{o3}(\overline{T}_{a3} - \overline{T}_{w3}) \\ (h_{in} - \overline{h}_l) \dot{m}_{in} + \pi D_i L_1 \alpha_{i1}(T_{w1} - \overline{T}_1) / (0.5 A \overline{\rho}_l (h_{in} - \overline{h}_l)) \end{bmatrix} \tag{8}$$

$$g\left(\mathbf{x}_r, \mathbf{y}_r, \mathbf{u}_e\right) = \begin{bmatrix} A\left(\bar{\rho}_1 h_l - \bar{\rho}_3 h_g\right)\left[\left(h_{in} - h_l\right)\dot{m}_{in} + \pi D_l L_1 \alpha_{i1}(T_{w1} - \bar{T}_1)\right] \\ -\dfrac{1}{2} A\bar{\rho}_1\left(h_{in} - h_l\right)\left[h_l \dot{m}_{in} - h_g \dot{m}_o + \pi D_l L_2 \alpha_{i2}(T_{w2} - \bar{T}_2)\right]; \\ \dfrac{1}{2} A\bar{\rho}_3\left(h_g - h_o\right)\left[\left(h_{in} - h_l\right)\dot{m}_{in} + \pi D_l L_1 \alpha_{i1}(T_{w1} - \bar{T}_1)\right] \\ -\dfrac{1}{2} A\bar{\rho}_1\left(h_{in} - h_l\right)\left[\left(h_g - h_o\right)\dot{m}_o + \pi D_l L_3 \alpha_3(T_{w3} - \bar{T}_3)\right]; \\ A\left(\bar{\rho}_1 - \bar{\rho}_3\right)\left[\left(h_{in} - h_l\right)\dot{m}_{in} + \pi D_l L_1 \alpha_{i1}(T_{w1} - \bar{T}_1)\right] \\ -\dfrac{1}{2} A\bar{\rho}_1\left(h_{in} - h_l\right)\left(\dot{m}_{in} - \dot{m}_o\right) \end{bmatrix} \quad (9)$$

The validity and accuracy of the above model have been verified in the literature [22] and in our previous work [12], demonstrating its ability to effectively capture the dynamic characteristics of the evaporator.

### 2.2.2. Expander and pump

The dynamic behavior of the ORC system is primarily determined by the evaporator while the dynamics of the remaining components, such as the expander and pump, are significantly faster. Therefore, these two models can be described using static algebraic equations. Specifically, a quasi-steady-state black-box modeling approach [25] is used to model these components in this study.

For the expander, we assume that the isentropic efficiency coefficient ($\eta_{x,is}$) and the volume efficiency coefficient ($\eta_{x,vol}$) are the constant, then the model can be described by the following equations:

$$\dot{m}_x = \eta_{x,vol} V_x \rho_{x,in} \dot{\omega}_x \quad (10)$$

$$h_{x,out} = h_{x,in} + \eta_{x,is}\left(h_{x,is} - h_{x,in}\right) \quad (11)$$

where $\dot{m}_x$ represents the mass flow rate of the working fluid at the expander's inlet, $\rho_{x,in}$ represents the inlet density of the organic fluid, $V_x$ denotes the expander's swept volume, $\dot{\omega}_x$ refers to the rotation speed, $h_{x,out}$ and $h_{x,in}$ represent the downstream and upstream enthalpies, respectively, and $h_{x,is}$ stands for the working fluid's isentropic enthalpy. Accordingly, the output power of the expander during power generation can be calculated by

$$\dot{W}_x = \left(h_{x,in} - h_{x,out}\right)\dot{m}_x \tag{12}$$

Similarly, the pump used in this paper has fixed displacement, so the mass flow rate through the pump can be modeled as

$$\dot{m}_p = \eta_{p,vol}V_p\rho_{p,in}\dot{\omega}_p \tag{13}$$

where $V_p$ represents the pump displacement, $\eta_{p,vol}$ signifies the volumetric efficiency of the pump, $\rho_{p,in}$ represents the density at the pump's inlet, and $\dot{\omega}_p$ denotes the rotation speed. The consumed power of the pump during the operation can be calculated by

$$\dot{W}_p = \dot{m}_p\frac{P_{p,out} - P_{p,in}}{\rho_{p,in}\eta_{p,is}} \tag{14}$$

where $P_{p,out}$ and $P_{p,in}$ represent the outlet and inlet pressures, respectively, and $\eta_{p,is}$ signifies the coefficient of isentropic efficiency of the pump.

## 2.3. Control-oriented design & analysis

Regarding the control of ORC systems for waste heat recovery, many research papers mainly focused on the control of the SH at the outlet of the evaporator [19], which is one of the most important parameters and indicators for the operation of the system. However, in actual industrial production, the waste heat working conditions generated by the factory often change, resulting in sudden changes in the mass flow rate and temperature of the waste heat. When the ORC waste heat recovery system operates in follow-the-energy (FTE) mode, the speed of the expander usually varies with the heat source. At this time, the system must maintain the outlet temperature (or SH) of the evaporator in the presence of uncertain external disturbances [9].

Therefore, this paper focuses on the setpoint tracking control of $SH$, which serves as the control objective, and the pump speed $\omega_p$ is considered as the manipulated variable. Since the ORC system needs to maintain a user-defined time-varying degree of superheat, the disturbance variable of the control system would be the operating condition of the waste heat (i.e., the waste heat mass flow rate $m_{al}$).

## 2.4.   Reinforcement learning for ORC superheat control

As mentioned in the introduction section (Section 1), PID and MPC are commonly used and widely accepted control algorithms for ORC superheat control. However, MPC may not perform well in uncertain scenarios and is sensitive to model-plant mismatch. On the other hand, DRL has emerged as a promising alternative because it does not require explicit knowledge of the specific model of the controlled system and learns through trial and error by interacting with the environment. Additionally, the inherent random exploration of DRL greatly enhances its robustness in uncertain environments. Recently, there have been many applications and improvements of DRL in the process control field [16-18, 26-29].

DRL is a learning paradigm that differs from traditional supervised and unsupervised learning. It adjusts its control strategy by continuously interacting with the environment $E$, aiming to maximize the cumulative reward. The agent receives a reward signal and adjusts its behavior accordingly. Typically, the Markov Decision Process (MDP) is used to model the RL task, and $\mathcal{S}$ stands for the state space and $\mathcal{A}$ is the action space. Within each time step, the RL agent executes an action $\mathbf{a}_t$ based on the current control policy $\pi_\theta(\mathbf{a}_t | \mathbf{s}_t)$ parameterized by $\theta$, receives a corresponding scalar reward signal (immediate reward) $r_t = r(\mathbf{s}_t, \mathbf{a}_t) : \mathcal{S} \times \mathcal{A} \mapsto \mathbb{R}$ from the environment, and then moves the state $\mathbf{s}_t$ to the next state $\mathbf{s}_{t+1}$.

Specifically, the episode reward $R_t$ can be calculated by the (discounted) sum of immediate rewards $r_t$, i.e.,

$$R_t = \sum_{t=1}^{T} \gamma^t r_t = \sum_{t=1}^{T} \gamma^t r(\mathbf{s}_t, \mathbf{a}_t). \tag{15}$$

Then, the RL objective will be defined as

$$J\left(\pi_\theta\right) = \max_{\pi_\theta} \mathbb{E}_{\mathbf{s} \sim \rho^\pi, \mathbf{a} \sim \pi_\theta} \left[ R_t | \mathbf{s}_t, \mathbf{a}_t \right]. \tag{16}$$

In DRL, a value function is introduced to evaluate the outcome of a given state or state-action pair. One common choice is the state-action value function [30], denoted as the expected value of the episode reward $R_t$ under the current policy $\pi$, i.e,

$$Q^\pi(\mathbf{s}_t, \mathbf{a}_t) = \mathbb{E}_\pi \left[ R_t | \mathbf{s}_t, \mathbf{a}_t \right] \tag{17}$$

The Bellman equation in optimal control is then iteratively used to solve the optimization problem stated in Eq. (16):

$$Q^{\pi}(\mathbf{s}_t, \mathbf{a}_t) = \mathbb{E}_{r_t, \mathbf{s}_{t+1} \sim E}\left[ r(\mathbf{s}_t, \mathbf{a}_t) + \gamma \mathbb{E}_{\mathbf{a}_{t+1} \sim \pi}\left[ Q^{\pi}(\mathbf{s}_{t+1}, \mathbf{a}_{t+1}) \right] \right]. \tag{18}$$

## 3. Methodology

### 3.1. Framework of DRL-based Sim2Real control design

The purpose of this paper is to propose a simple Sim2Real transfer learning method for the DRL-based controller of the ORC system, which addresses the issues of low sample efficiency, time consumption, and safety risks encountered in the interactive training with the ORC system.

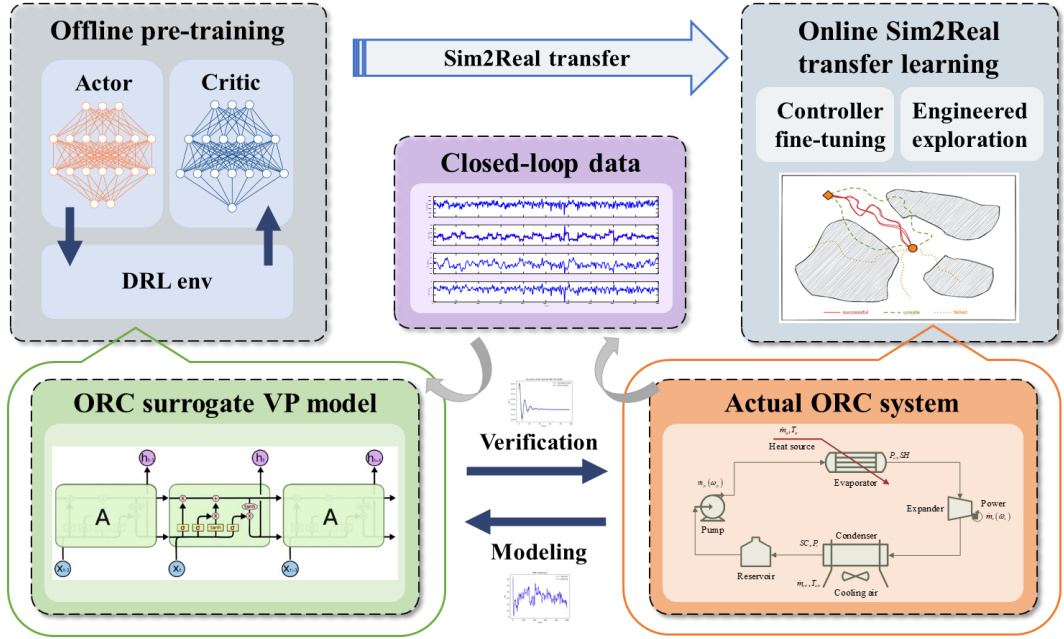

**Fig. 3    Framework of DRL-based Sim2Real control design.**

The overall framework of the algorithm is shown in Fig. 3. It can be divided into three stages:

1. **Surrogate modeling phase:** Historical closed-loop operational data is used to establish a surrogate model based on LSTM neural networks. After verifying its accuracy and precision, the surrogate model is used as a virtual prototype of the ORC system.

2. **Offline pre-training phase:** The DRL agent is trained using the surrogate model

in a simulation environment with strong fluctuations in the ORC waste heat conditions. This approach avoids direct interaction with the actual ORC system, thereby reducing safety risks.

3. **Online Sim2Real transfer phase:** Sim2Real transfer learning is used to address the model-plant mismatch between the surrogate-based virtual prototype and the physical ORC system.

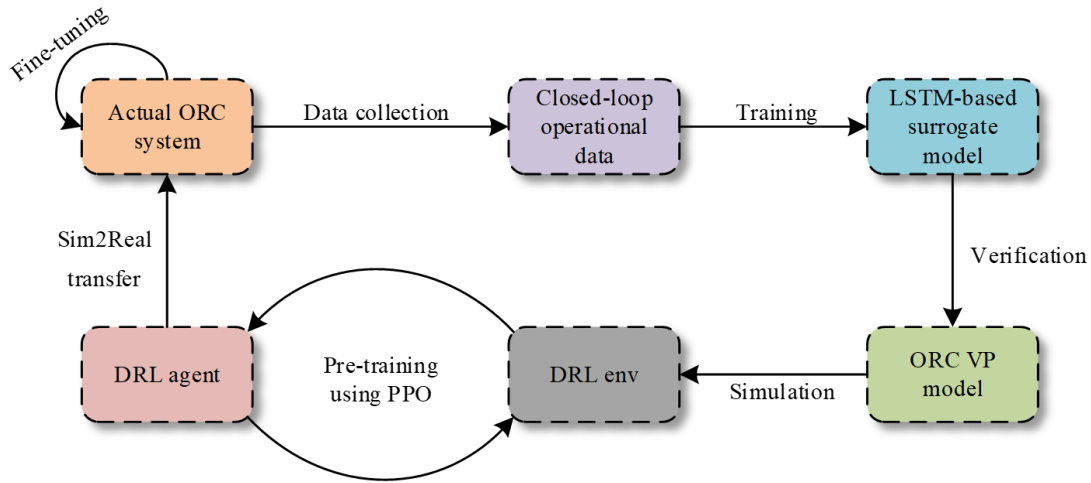

**Fig. 4    Workflow diagram of DRL-based Sim2Real control design.**

Specifically, a large amount of historical closed-loop operational data is collected from the actual ORC system under conditions of disturbance in waste heat and variations in superheat setpoints. The collected data is then used to establish a surrogate model of the ORC system based on LSTM recurrent neural networks. After verifying the accuracy of the trained surrogate model, it is used as the virtual prototype of the ORC system. The virtual prototype is employed as a pre-training simulation environment for DRL, within which the Proximal Policy Optimization (PPO) algorithm is selected as the RL agent for the interaction and training. This process allows the agent to learn an initial control policy. Subsequently, the RL agent that achieves satisfactory control performance in the simulation environment is selected and transferred to the actual ORC system for Sim2Real adaptation of the controller. Through fine-tuning, the agent is further trained to achieve the same desired control performance on the real system. Therefore, the proposed method allows for rapid control design based on DRL while effectively mitigating the safety risks associated with direct online training. The workflow diagram and algorithm flowchart of the proposed method are shown in Fig. 4 and Fig. 5, respectively.

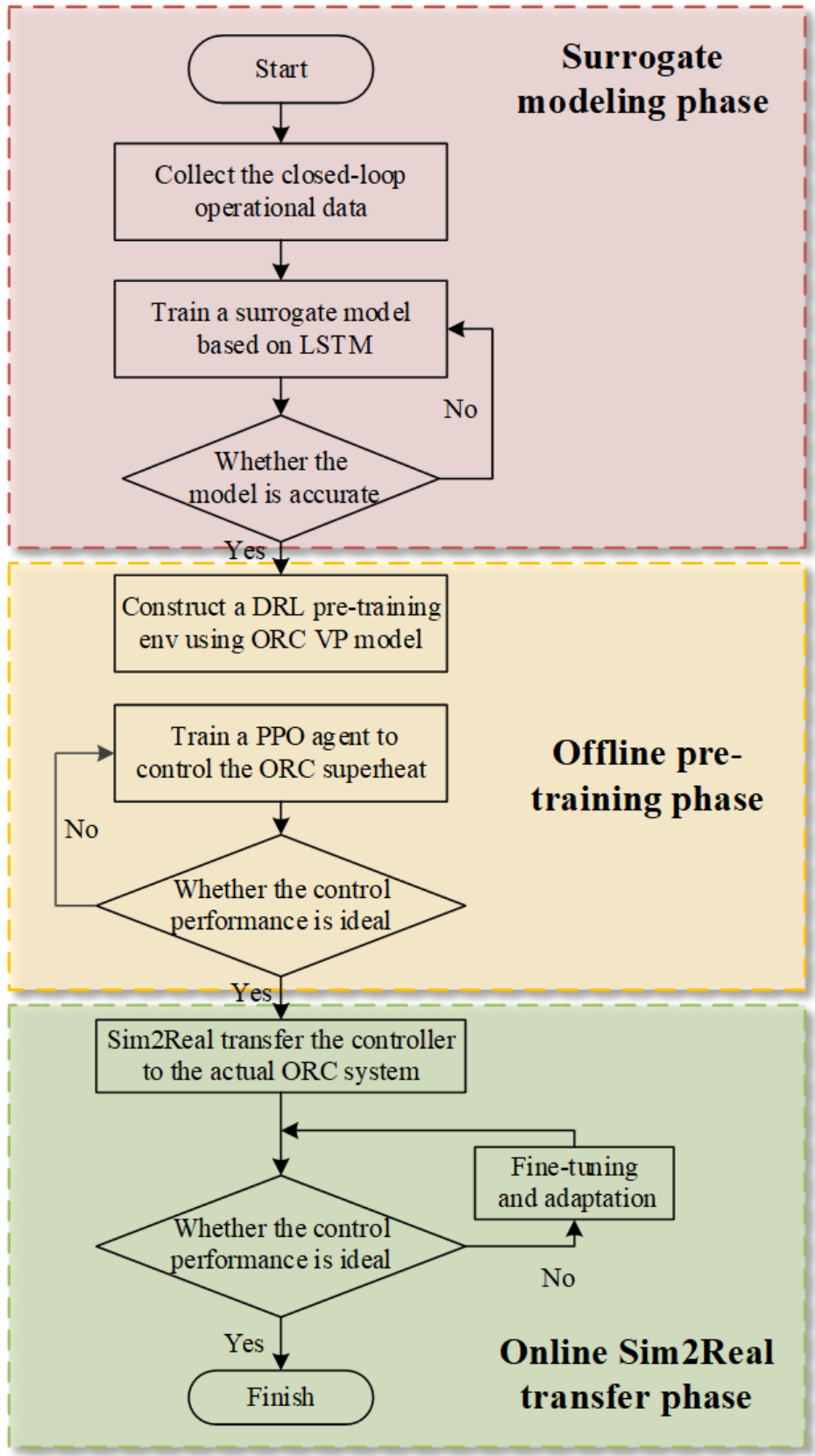

**Fig. 5 Flowchart of the DRL Sim2Real transfer learning algorithm.**

## 3.2. Surrogate modeling for virtual ORC prototype

### 3.2.1. Motivation for building an ORC surrogate model

Based on the framework presented in Section 3.1, Sim2Real serves as a bridge between the simulation environment and the real ORC system. In the offline simulation/training phase, it is necessary to construct a virtual prototype model of the ORC system as an environment for the DRL agent. Virtual prototyping [31] uses a digital model, known as a virtual prototype (VP), to assess and evaluate the specific attributes of a product or manufacturing process within a computational environment, thereby replacing the need for a physical prototype.

In this paper, a VP model tailored for ORC superheat control is constructed using surrogate modeling techniques commonly used in process optimization and control. This allows us to train the RL agent in a simulation environment to achieve the desired control performance. However, it should be noted that the dynamic characteristics of the VP model may differ from those of the real system, necessitating subsequent Sim2Real transfer learning.

It is worth mentioning that the historical closed-loop operational data of the ORC system is used to build the surrogate-based VP model in this study. On the one hand, typical waste heat recovery ORC systems have accumulated a wealth of closed-loop operational data during their production processes. Utilizing this historical data eliminates the need to perform conventional identification tests, saving time and labor costs by avoiding system shutdowns for separate identification tests. On the other hand, the closed-loop operational data is generated by the system's existing controller (e.g., PID), making it well-suited to serve as the environment for DRL interaction. Since the RL agent's optimization decisions follow the framework of an MDP, an ORC surrogate model constructed from closed-loop data is advantageous for control-oriented modeling and design.

### 3.2.2. Long Short-Term Memory (LSTM) networks

The historical operating data of the ORC system contains key information such as the system's states at different time steps and the transition relationships of states over time under control signals. However, due to the strong coupling and nonlinearity of the system, traditional algorithms such as linear regression and PLS struggle to ensure accuracy when using historical operating data to build a surrogate model of the ORC system. In contrast, the LSTM model is known for its ability to address nonlinear modeling problems. LSTM is a variant of recurrent neural networks (RNNs) that has shown excellent performance in handling sequence data tasks. An LSTM cell consists of input gates, forget gates, output gates, and a cell state, responsible for processing and propagating input data. The specific structure is depicted in Fig. 6.

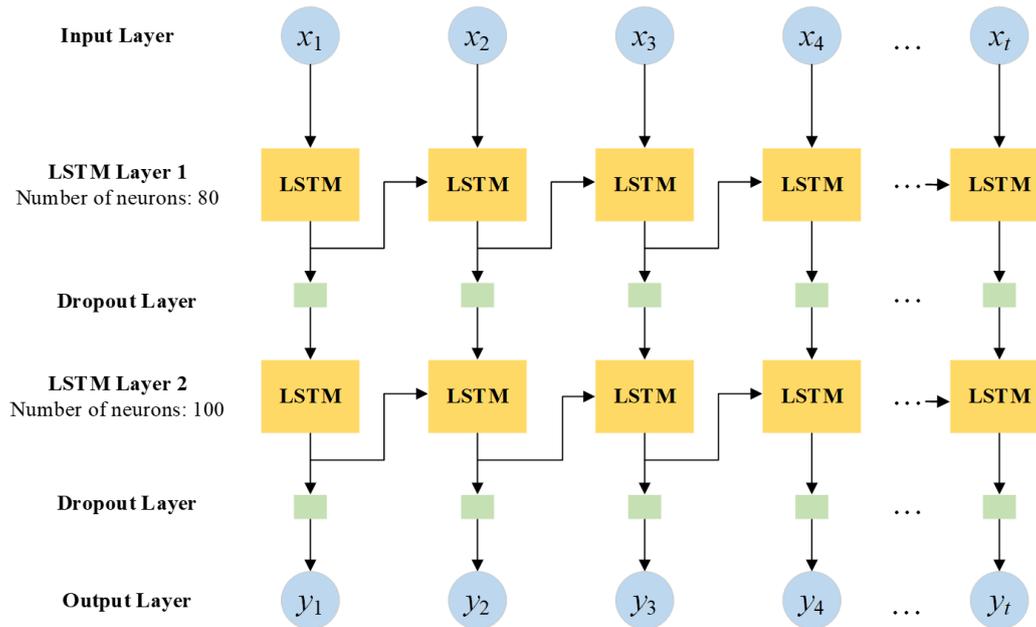

**Fig. 6    Structure diagram of the LSTM-based ORC surrogate prediction model.**

Compared to traditional algorithms, the LSTM model offers the following advantages:

1.  LSTM can capture long-term dependencies within sequences, making it effective in handling time-series data of ORC system state variables that exhibit temporal variations.

2.  LSTM can learn and model nonlinear relationships, whereas traditional algorithms such as linear regression and PLS often assume linear relationships among the data. This allows LSTM to better adapt to the complex data patterns and nonlinear relationships present in the ORC system.

3.  Through the cell state and gate mechanisms, LSTM can effectively remember and use information from past time steps to better handle tasks that require consideration of long-term dependencies.

In summary, LSTM can fully exploit the rich information contained in the historical operating data of the ORC system. Selecting LSTM as a surrogate model for the ORC system helps minimize the differences between the model and the physical system, thus ensuring modeling accuracy. During testing, the output state of the previous time step serves as the input state for the next time step, and the predictive information from all previous states is continuously propagated and combined with the input state to predict the output state of the next time step. Specifically, when training the LSTM-based ORC surrogate model, the input variables $x_i (i = 1, 2, \cdots, t)$ will include the RL state variables $s_t$ (i.e., the evaporator outlet pressure $P_e$, the evaporator outlet temperature $T_{oe}$, the mass flow rate of the working fluid $m_{ai}$, the speed of the working fluid pump $\omega_p$, the superheat value $SH$, and the tracking error $SH - SH_{set}$) and RL action variables $a_t$ (i.e., the control signals $\omega_p$); and the input variables $x_i (i = 1, 2, \cdots, t)$ will include the RL state variables $s_{t+1}$ of the next time step.

## 3.3. Learning initialized controller by pre-training in surrogate environment

### 3.3.1. Observation, action, and reward

As discussed in Sections 2.3 and 2.4, the objective of the ORC DRL control in this paper is to maintain the SH at its setpoint. The setpoint may vary over time, and there may be significant disturbances in the system, such as fluctuations in the heat source flow rate caused by the operation of plant equipment. Given the need for setpoint tracking, the tracking error of the controller is included as one of the states in the DRL framework, effectively incorporating the setpoint information. The other variables in the DRL states represent key state parameters of the system, including the mass flow rate of the waste heat. For ease of computation, the range of the DRL action is constrained between -1 and 1, which is equivalent to normalizing the data.

Thus, in the design of the RL agent for ORC superheat control, the RL state is formulated as $\mathcal{S} \triangleq \begin{bmatrix} P_e, T_{oe}, m_{ai}, \omega_p, SH, e \end{bmatrix}^T$, which includes the pressure and the temperature at the evaporator's outlet, the waste heat's mass flow rate, the pump speed, SH, and the error signal, respectively. On the other hand, the RL action $\mathcal{A} \triangleq \omega_p$ is defined as the control input, specifically the normalized pump speed $\omega_p$, which is mapped within the range [-1, 1].

In the framework of this study, the reward function is divided into continuous reward and discrete reward, considering that providing a mixed reward signal would be beneficial in many cases for the design of setpoint tracking controllers. The reason for this is continuous reward signals can generally improve convergence during training while discrete reward signals may lead the agent to better rewarding regions in the state space of the environment. Specifically, the total reward $r_t$ is calculated as the sum of continuous rewards and discrete rewards, i.e.,

$$r_t = r_{1,t} + r_{2,t} \tag{19}$$

where $r_{1,t}$ is the continuous part of the reward,

$$r_{1,t} = -0.8 \times e^2(t) - 0.8 \times |u(t) - u(t-1)| \tag{20}$$

and $r_{2,t}$ is the discrete reward in Table 1, which is proportional to the absolute value of the error signal. Therefore, the discrete reward signal is used to drive the ORC superheat control system away from undesirable or unsafe states and towards the setpoint, and then the continuous reward signal is added to improve convergence by providing a smooth reward near the SH setpoint.

**Table 1 Discrete reward setting in the experiments**

| $|e(t)|$ | $< 0.05$ | $< 0.1$ | $< 0.5$ | $< 1$ | $< 2$ | $< 3.5$ | $< 5$ | $\geq 5$ |
|---|---|---|---|---|---|---|---|---|
| $r_{2,t}$ | 300 | 100 | 50 | 0 | -5 | -20 | -50 | -100 |

### 3.3.2. DRL agent & algorithm setting

1) Proximal Policy Optimization (PPO) agent

Proximal Policy Optimization (PPO) is a classical DRL algorithm [32] that has been introduced in recent years and is widely recognized as one of the best performing DRL algorithms. In this paper, PPO is chosen as the training algorithm for ORC superheat control. PPO is a model-free, online, on-policy, policy gradient RL method that involves two main steps: sampling data through environmental interaction and optimizing a clipped surrogate objective function using stochastic gradient descent [32]. PPO addresses the problem that policy gradient algorithms are sensitive to step size, as large differences between the old and new policies can hinder learning. PPO introduces a new objective function that allows for small-batch updates over multiple training steps, thus solving the challenge of determining appropriate step sizes in policy gradient algorithms.

For PPO-Clip, the objective is to update policies via

$$\theta_{k+1} = \arg\max_{\theta} \mathop{\mathbb{E}}_{s,a \sim \pi_{\theta_k}} \left[ L(s,a,\theta_k,\theta) \right], \tag{21}$$

where the term $L$ is defined as:

$$L(s,a,\theta_k,\theta) = \min\left( \frac{\pi_{\theta}(a\,|\,s)}{\pi_{\theta_k}(a\,|\,s)} A^{\pi_{\theta_k}}(s,a), \; \text{clip}\left( \frac{\pi_{\theta}(a\,|\,s)}{\pi_{\theta_k}(a\,|\,s)}, 1-\epsilon, 1+\epsilon \right) A^{\pi_{\theta_k}}(s,a) \right), \tag{22}$$

where $\theta_k$ represents the policy parameters before the update (at iteration $k$), and $\epsilon$ is a small hyperparameter that determines the allowable deviation of the new policy from the old one. And the importance sampling ratio, $\pi_{\theta}(a\,|\,s) / \pi_{\theta_k}(a\,|\,s)$, allows sample reuse in the on-policy algorithm.

By limiting the range of policy changes with the clipping operation described in Eq.(22), PPO-Clip encourages policies to stay close to the existing old policy. Typically, multiple steps of minibatch stochastic gradient descent are performed to maximize the objective. There is another simplified version of the surrogate objective function that depends on the advantage function:

$$L(s,a,\theta_k,\theta) = \min\left( \frac{\pi_{\theta}(a\,|\,s)}{\pi_{\theta_k}(a\,|\,s)} A^{\pi_{\theta_k}}(s,a), \; g(\epsilon, A^{\pi_{\theta_k}}(s,a)) \right), \tag{23}$$

where

$$g(\epsilon, A) = \begin{cases} (1+\epsilon)A & A \geq 0 \\ (1-\epsilon)A & A < 0. \end{cases} \tag{24}$$

As a result, the clipping operation acts as a form of regularization by reducing the incentives for significant policy changes, and the hyperparameter $\epsilon$ determines the allowable degree of deviation from the old policy while still producing favorable outcomes.

2) Algorithm setting

The critic network in the RL agent architecture consists of two paths: the state path and the action path, along with a common path. Each of these paths contains hidden layers with specific neuron configurations and activation functions. Specifically, both the state and the action paths consist of a hidden layer with 300 neurons. The common path, responsible for $Q$-value estimation, is designed with a layer of 200 neurons.

They are both activated by the ReLU function.

The actor network consists of two hidden layers. The first hidden layer consists of 300 neurons, while the second contains 200 neurons. The output layer of the actor network corresponds to the action dimension. Both layers are activated by the tanh activation function, which maps the output to a range between -1 and 1, ensuring bounded and smooth action values for the actor network.

During the training process, the action $a_t$ generated by the actor network, referred to as the real-time control signal (i.e., the pump speed $\omega_p$), is directly used and combined with the current state $s_t$ as inputs to the surrogate model, which predicts the state $s'_{t+1}$ at the next time step. The PPO agent then takes a new action $a_{t+1}$ based on the observed state $s_{t+1}$ and the computed reward $r_t$ to iteratively update the control policy. The hyperparameters used for pre-training with the PPO algorithm are listed in Table 2.

**Table 2 Hyperparameters of the PPO algorithm during training.**

| Hyperparameters | Value |
|---|---|
| critic learning rate | 3e-4 |
| actor learning rate | 2e-4 |
| n_epoch | 10 |
| discount factor | 0.9 |
| lam | 0.95 |
| entropy coef | 0 |
| minibatch size | 64 |
| replay buffer capacity | 2e6 |

Note that the specific values of the hyperparameters will depend on the experimental setup and can be adjusted through experimentation and tuning to achieve the best performance for the ORC system control task.

### 3.3.3. Pre-training under large-scale variation of waste heat working conditions

Unlike traditional pre-training objectives that aim for high reward performance within a specific operating range, the approach in this study emphasizes the average performance of the surrogate-based pre-training in scenarios with large variations of waste heat operating conditions. Achieving optimal control performance under specific fixed waste heat conditions is less concerned and instead the robustness of the learned control policies in the presence of significant fluctuations is focused on. This emphasis on stochasticity is particularly beneficial for transfer learning. If the agent can achieve satisfactory but not necessarily optimal results in various complex scenarios within the surrogate environment (VP model), it is more likely to learn robust and adaptive control strategies in transfer learning settings. Therefore, the pre-training in this study should be conducted under conditions of large variation in waste heat operating conditions to provide a solid foundation for the subsequent Sim2Real transfer learning.

## 3.4. Sim2Real transfer of ORC superheat control

### 3.4.1. Sim2Real transfer to alleviate model-plant mismatch

Due to the inherent deviations between the simulation environment and the actual ORC system, directly using the DRL controller trained on a surrogate model as the control policy for the actual system introduces the risk of model-plant mismatch, which can significantly degrade the control system performance. In extreme cases, significant model-plant mismatch can even lead to unsafe behaviors of the DRL-based ORC controller. This is one of the key challenges hindering the practical application of DRL in industrial settings. DRL agents need to interact with the environment, but they are often sensitive to variations in environmental dynamics. Since it is practically impossible to obtain a perfect simulator, the issue of model-plant mismatch must be carefully considered.

Therefore, this paper adopts the concept of Sim2Real transfer from transfer learning to mitigate the risks of model-plant mismatch and enhance the control performance of the DRL controller from the simulation environment to the real system. First, the pre-training of the DRL controller is conducted using the constructed surrogate model to achieve satisfactory control performance in the simulation environment. It is important to note that the waste heat operating conditions in the simulation environment undergo significant variations, which helps the agent improve its robustness and adaptability. The pre-trained initial controller is then validated in the real ORC system, where the control performance may not be as good as in the simulation. However, this is not a concern as the concept of Sim2Real transfer allows for fine-tuning in real-world scenarios. Leveraging the partially effective initial controller as a baseline, Sim2Real transfer allows the agent to appropriately explore and learn about the actual ORC system. Over time, as online training progresses, the degree of model-plant mismatch diminishes, and then the training is terminated, allowing the controller to be deployed for real-world online operation.

### 3.4.2. Engineered level of exploration for better fine-tuning and safety

It is well known that a crucial aspect of DRL is the trade-off between exploration and exploitation. The agent needs to balance between exploiting known optimal actions and exploring unknown ones. Overemphasis on exploitation can cause the agent to get stuck in local optima, hindering its ability to explore the environment, and potentially missing out on better strategies. Conversely, over-exploration can waste time and resources, prolonging convergence. During training, the agent inevitably explores unsafe actions, which poses significant risks in process control domains such as the chemical industry. Inappropriate exploration can result in serious consequences, including equipment damage, environmental pollution, and threats to human life. Therefore, when applying DRL to industrial control scenarios, it is crucial to address the exploration-exploitation dilemma and adjust the level of exploration appropriately.

In the context of Sim2Real transfer learning for fine-tuning and adaptation of DRL controllers, carefully designed or engineered exploration can lead to better fine-tuning performance while inappropriate exploration can deteriorate the performance of the pre-trained controller in the surrogate environment. Since the PPO algorithm is used as a DRL controller, where exploration is primarily achieved through the variance part of the actor's output. A larger variance represents more intensive exploration since the actions fed into the environment are sampled from the mean and variance output of the actor network. Intuitively, it is necessary to limit excessive exploration to avoid undermining the achievements of pre-training. This is because, at times, when the agent is operating outside the current local region, it can easily get stuck in another local region and fail to return to the original one, thereby nullifying the effects of pre-training. However, the exploration should not be too small either, as extremely small exploration fails to address the model-plant mismatch issues. Therefore, in the setting of Sim2Real transfer learning, compared to the pre-training phase, the level of exploration should be appropriately reduced during fine-tuning to prevent the agent from deviating from the action distribution range obtained from pre-training, which could cause a decline in controller performance, violation of constraints, or unsafe operations.

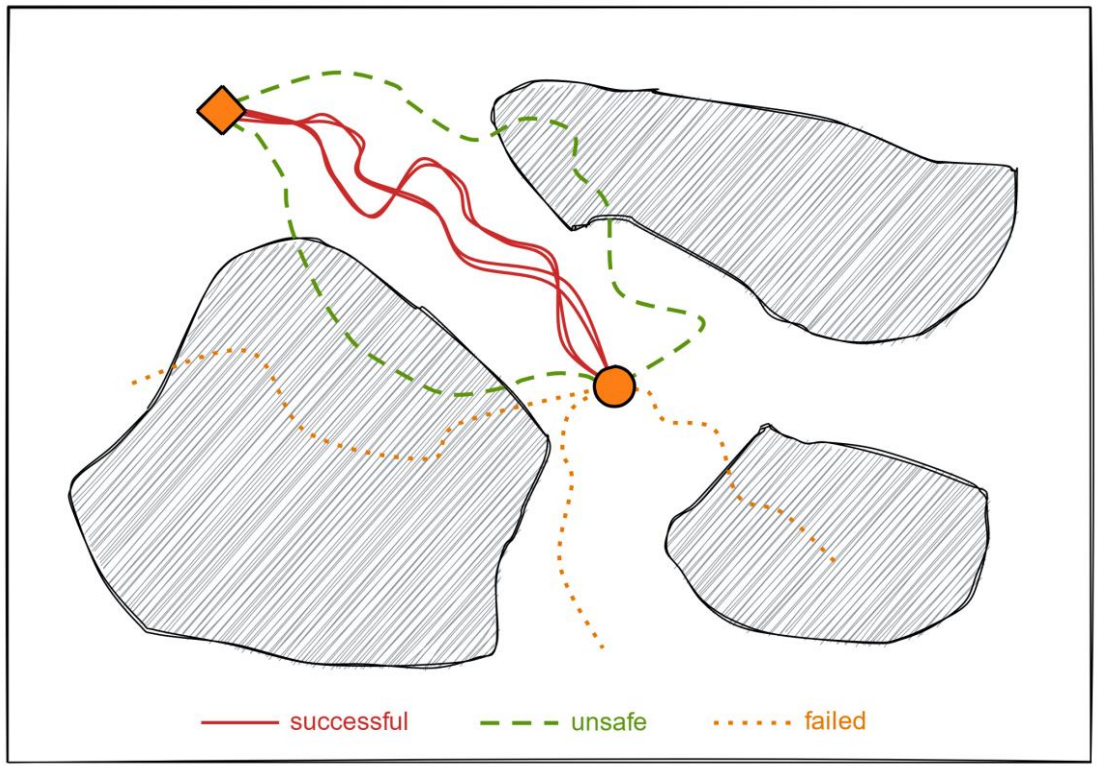

**Fig. 7 Schematic diagram illustrating the importance of the level of exploration.** The circle represents the initial position of the agent, the diamond represents the final goal that needs to be achieved, and the shaded area represents the unsafe region in the state and action spaces. The red solid line is the trajectory that reaches the goal successfully without unsafe exploration, the green dashed line is the unsafe trajectory, and the orange dotted line is the failed attempt.

Fig. 7 visually depicts the impact of exploration on Sim2Real transfer learning in the DRL state space. The red trajectory demonstrates the successful and safe achievement of the desired goal through carefully designed exploration, achieving Sim2Real transfer with fewer episodes. The green trajectory illustrates the consequences of excessive exploration, which violates constraints or enters unsafe operating regions. The orange trajectory represents the detrimental effects of inappropriate exploration, which destroys the fruitful results of pre-training and leads to training failure and collapse. Thus, for Sim2Real scenarios, the level of exploration is a crucial aspect of transfer learning that needs to be carefully designed and treated as a key hyperparameter for fine-tuning. In the next section, the importance of the exploration variance will be specifically demonstrated.

## 4. Results & discussion

In this section, the methodology part is demonstrated through experiments. Based on the whole design scheme discussed in previous sections, the first step is to learn an agent with PPO using ORC VP model; then the experimental results of Sim2Real transfer learning are followed, including the discussions on the training efficiency, control performance, and sensitivity analysis.

### 4.1. Closed-loop data-driven surrogate modeling

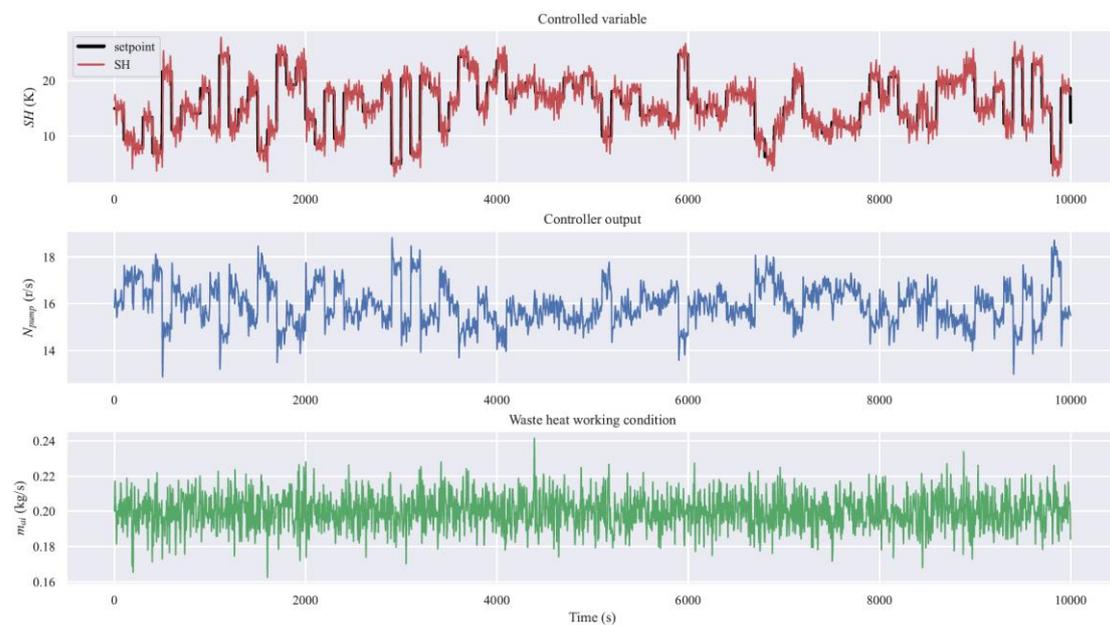

**Fig. 8 Closed-loop operational data acquisition of the waste heat recovery ORC system under large-scale variation of working conditions and disturbances.**

To establish a good surrogate model, it is necessary to collect operating data from the actual ORC system. Specifically, the parameters of the PI controller are set as $P = 0.15, I = 0.03$. The SH of the ORC system is closed-loop controlled. During this period, the SH setpoint $SH_{set}$ is randomly changed every 100 seconds, and Gaussian white noise is added to the waste heat disturbance $m_{ai}$ to cover a wide range of operating conditions. A total of 10,000 historical data points during the closed-loop operation are collected. The random setpoint changes and the wide fluctuation of waste heat conditions are intended to make the closed-loop data more representative and rich, capturing the main dynamic characteristics of the ORC system. It is also intended to address the issue of process dynamic behavior being masked by closed-loop control. The collected data includes the system state variables $s_t$ and control signals $\omega_p$. The state variables consist of the evaporator outlet pressure $P_e$, evaporator outlet temperature $T_{oe}$, the mass flow rate of the working fluid $m_{ai}$, speed of the working fluid pump $\omega_p$, superheat value $SH$, and tracking error $SH - SH_{set}$. The results of the collected closed-loop operational data are shown in Fig. 8.

Next, a neural network prediction model should be trained. The LSTM network is selected as the surrogate model for the ORC system. The 10,000 collected operational data points are divided into two sets: the first 8,000 data points are used as the training set, and the remaining 2,000 data points are used as the test set. The ORC surrogate model is constructed based on the LSTM neural network. The first layer is an LSTM layer with 80 neurons, followed by a dropout layer to prevent overfitting. Another LSTM layer of 100 neurons is added, followed by another dropout layer to prevent overfitting. The state $s_t$ and control signal $\omega_p$ at the time step $t$ are fed into the neural network as inputs, and the state $s_{t+1}$ at time step $t+1$ is used as the label to train the LSTM prediction model. The results after training are shown in Fig. 9.

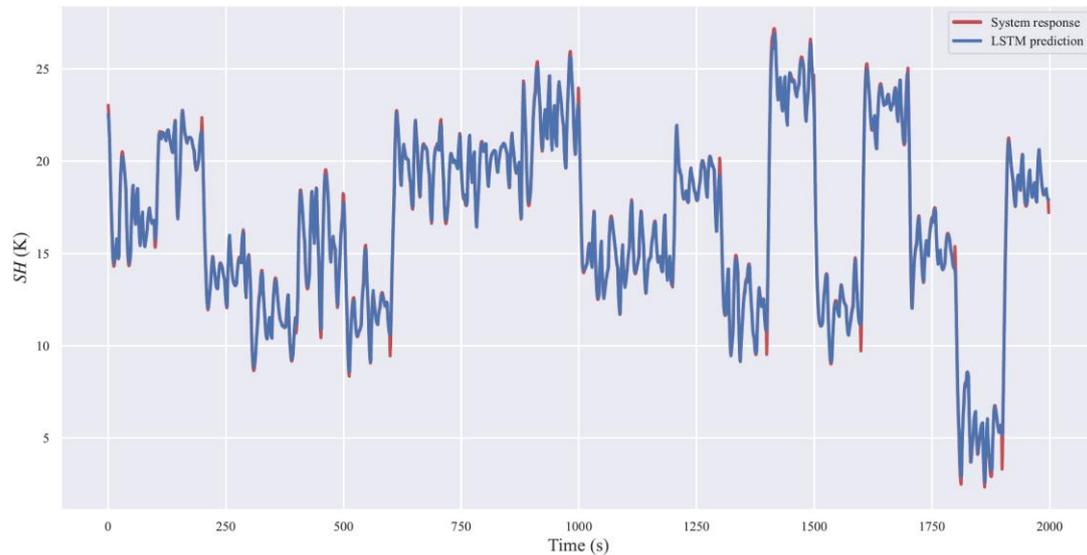

**Fig. 9 Prediction performance on the test set of the LSTM-based surrogate model.**

After calculation, the mean squared error (MSE) of the LSTM-based ORC surrogate model on the test set is determined to be 0.026519. To further validate the accuracy of the model, the same trajectory of the control signal (obtained from the PI control on the actual system) is assigned to both the trained ORC VP model and the ORC waste heat recovery system used for data collection. The two systems are then compared in terms of their output operation results while starting from the same initial state. The output operation results of the two systems are shown in Fig. 10. This approach allows a better test of the model accuracy under closed-loop control conditions. It can be observed that the LSTM-based ORC surrogate model exhibits high accuracy, which makes it suitable as the ORC VP model in this paper.

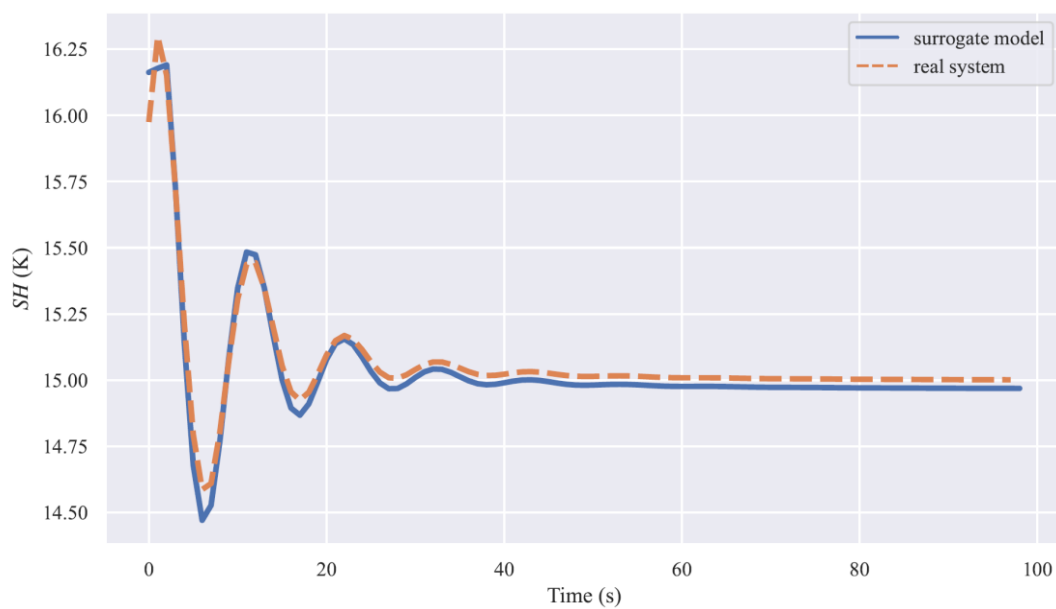

**Fig. 10 Use the same control sequence to verify the accuracy of the learned ORC surrogate model.**

### 4.2.  Model-free pre-training of PPO agent using learned ORC surrogate

The DRL environment based on the ORC surrogate model used in this experiment is built using OpenAI Gym[1], taking advantage of the compatibility between the Gym platform and various DRL algorithms. By embedding the constructed LSTM-based surrogate model in Gym, a user-defined simulation environment is obtained for interactive training of the DRL agent.

As described in Section 3.3, this study uses the PPO algorithm as the DRL controller for training. Based on OpenAI Gym, the surrogate model is used as the pre-training environment, and the PPO algorithm is used for learning, with the addition of time-varying waste heat disturbances. Specifically, to enhance the robustness and generalization of the DRL controller, a random trajectory of waste heat disturbance is generated in each episode, consistent with actual operating conditions so that the agent can suppress waste heat interference. Two training modes are set: one with a maximum episode length of 200 steps and a fixed single setpoint, and the other with a maximum episode length of 500 steps with setpoint changes occurring every 100 steps. Through multiple training iterations and parameter tuning, the rewards in both cases steadily increased, as shown in Fig. 11.

---

[1]  https://www.gymlibrary.dev/

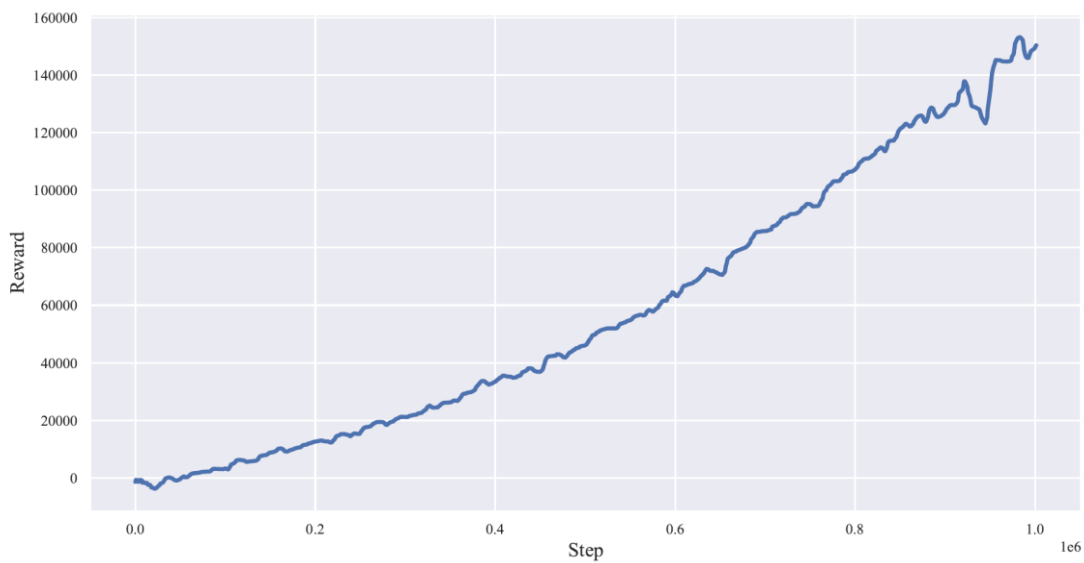

**Fig. 11 The pre-training process of our PPO-based ORC superheat controller.**

Subsequently, the pre-trained DRL controller is subjected to a performance test in the environment of the surrogate model. The results are presented in Fig. 12, which shows that the DRL controller, pre-trained on the surrogate model, exhibits the ability to rapidly reduce the tracking error to the setpoint even under significant variations in the waste heat disturbances, and outperforms the traditional PI control in various operating conditions. Thus, to date, a satisfactory initial control strategy has been successfully pre-trained under wide variations in operating conditions.

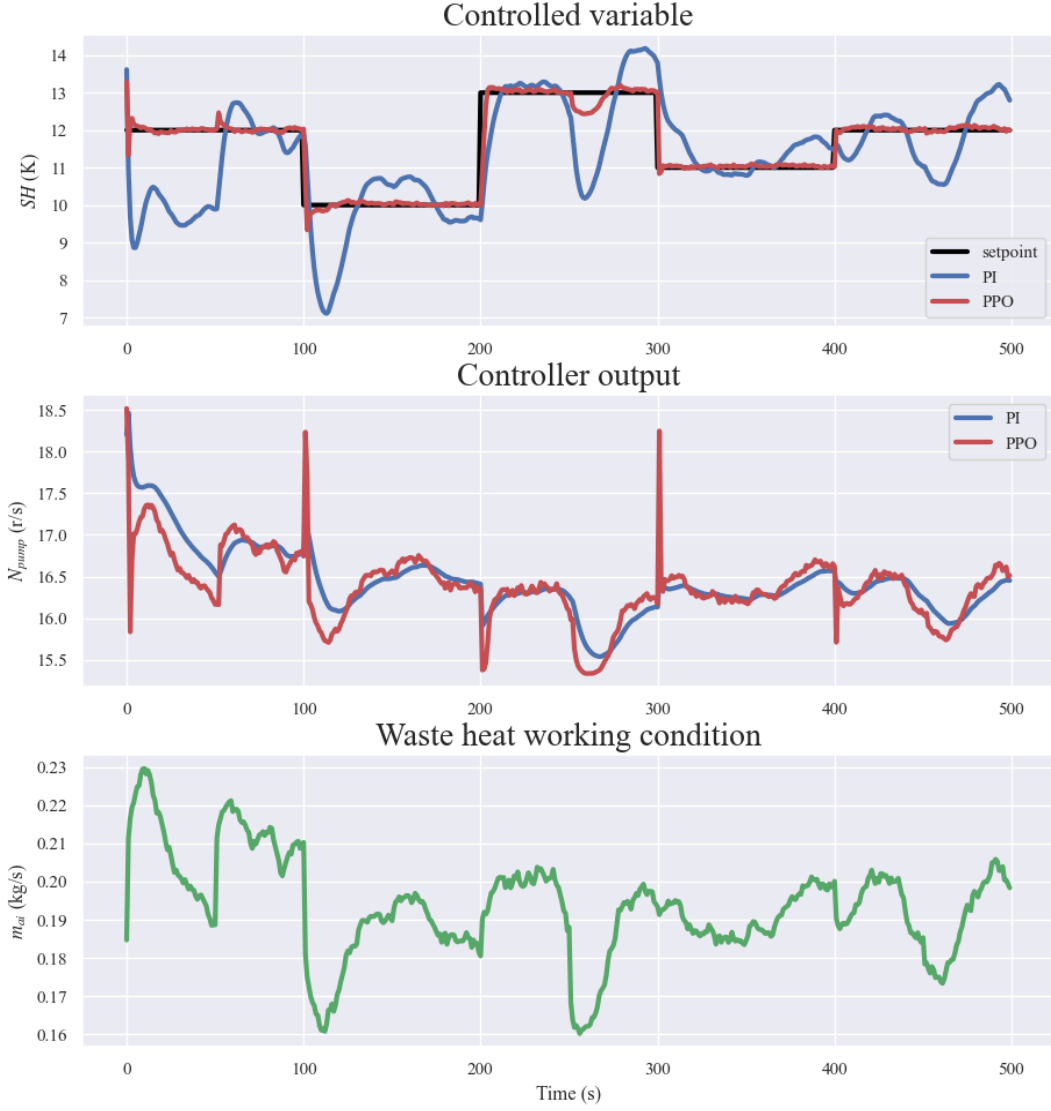

**Fig. 12 Control performances of the well pre-trained ORC superheat controller v.s. well-tuned PI controller in the ORC VP environment.**

## 4.3. Comparison of training rewards between Sim2Real transfer learning and directly learning from scratch

The aforementioned pre-training of the DRL controller based on the surrogate model gave satisfactory results. To show the advantages of Sim2Real transfer learning in the actual ORC system, two experiments are performed: 1) training the PPO agent directly from scratch, and 2) training it using our Sim2Real transfer learning method. Sim2Real transfer learning uses the pre-trained initial controller in the ORC VP environment as a benchmark for fine-tuning the actual waste heat recovery ORC system. On the other hand, directly learning from scratch refers to training the PPO controller directly on the real system. To ensure fairness, all hyperparameters are kept consistent between the two sets of experiments, except for whether the pre-trained actor network is loaded.

For each set of experiments, three different random seeds are used to test their average performance. The reward variations are compared between the proposed Sim2Real transfer DRL controller and the DRL controller trained directly from scratch, as shown in Fig. 13. In the figure, the solid lines represent the average rewards obtained from running the experiments multiple times, while the shaded regions represent the corresponding reward variances. A higher mean value indicates better control performance while a lower variance reflects better stability and robustness of the algorithm. According to the experimental results, it can be observed that the proposed Sim2Real transfer-based DRL controller achieves high rewards within a few episodes, indicating its effective utilization of the initialization from pre-training. On the other hand, the case of learning from scratch requires thousands of episodes to reach satisfactory rewards. Furthermore, our method demonstrates superior control performance compared to learning from scratch, as evidenced by the higher final reward, which is also supported by the results of the control performance comparison in Section 4.4.

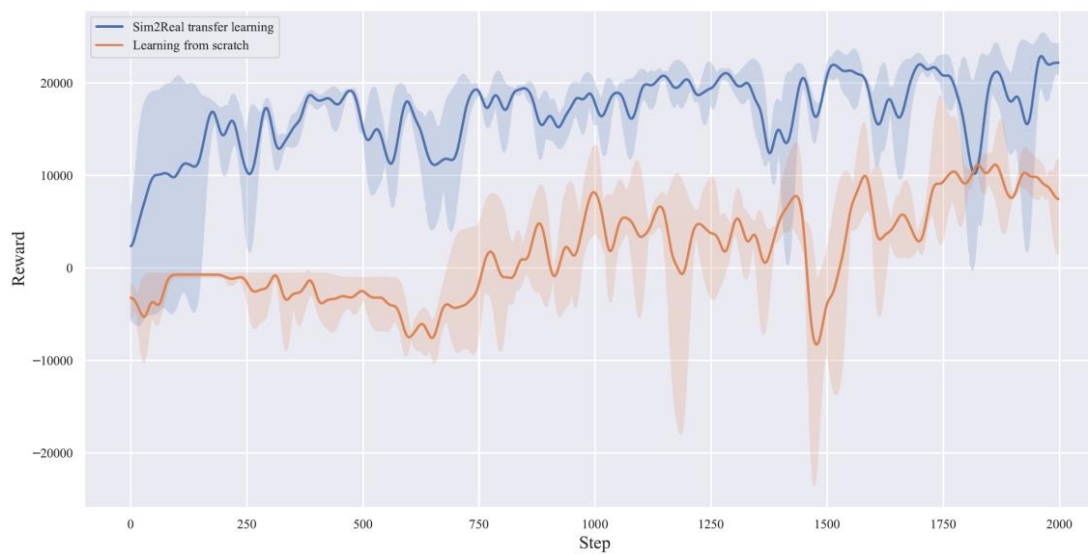

**Fig. 13 Comparison between Sim2Real transfer learning and directly learning from scratch. The results are obtained by replacing the random seeds by running 3 experiments at random to verify the robustness of the algorithm.**

## 4.4. Comparison of control performances between the DRL-based controller and conventional PI controller

In addition to evaluating the Sim2Real transfer learning performance on the actual ORC system, another comparative analysis between the proposed DRL controller and the traditional PI controller is performed in terms of control performance. During this test, the SH setpoint is randomly changed every 100 seconds to verify the control performance under different operating conditions. Besides, the waste heat mass flow rate is subjected to significant fluctuations to test the controller's disturbance rejection capability. It should be noted that the waste heat operating conditions applied to the actual ORC system are not encountered during the pre-training process, which better evaluates the effectiveness of Sim2Real transfer learning.

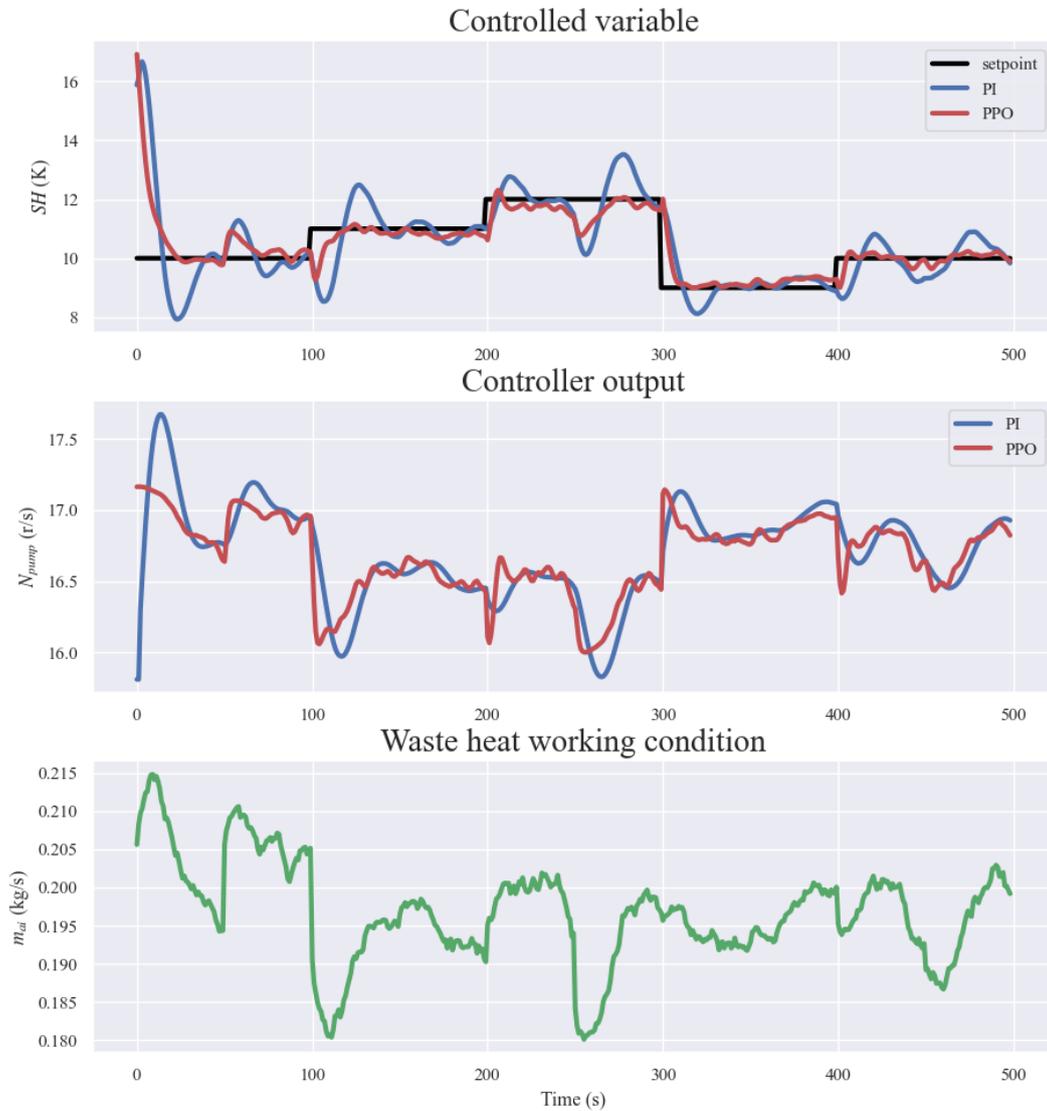

**Fig. 14   Control performances of the successfully trained (Sim2Real) DRL-based controller v.s. well-tuned PI controller under drastic changes in waste heat conditions.**

Based on the control results shown in Fig. 14, it can be observed that the successfully trained DRL-based controller outperforms the traditional PI control. The DRL controller achieves almost zero overshoot in tracking the setpoint while exhibiting a fast response. This is because the effect of the successfully trained DRL controller should be close to model-based optimized control approaches [17, 18], while the PI control requires adjustment based on the error feedback signal from the system output, which fundamentally cannot achieve fast tracking and zero overshoot. Importantly, the DRL controller exhibits significantly better disturbance rejection and maintains satisfactory control performance even in scenarios with drastic variations of waste heat in the ORC system. Although the PI controller exhibits strong robustness, this kind of robustness is achieved through passive adjustments based on its own closed-loop feedback. In contrast, the DRL controller actively suppresses disturbances based on the waste heat operating conditions of the ORC system, enabling it to quickly mitigate the impact of drastic variations in waste heat conditions on control stability.

### 4.5. Sensitivity analysis on the role of exploration variance

According to the analysis in Section 3.4.2, the exploration variance of the PPO agent is crucial for the Sim2Real transfer learning of the DRL-based controller. In the experimental design, the impact of the exploration variance of the actor network in the PPO algorithm is specifically considered during the training process. For this purpose, a simple sensitivity analysis is performed by varying the exploration variances while keeping other hyperparameters constant to test the performance of Sim2Real transfer learning.

These sensitivity analysis experiments are based on the same well pre-trained DRL controller for fine-tuning and adaptation. The initialized agent obtained in Section 4.2 is used and then transferred to the actual ORC system for Sim2Real transfer. Four sets of experiments are designed, each with a different exploration variance: 0.05, 0.1, 0.2, and 0.35. For each set of experiments, three different random seeds are used to test their average performance. The solid lines represent the means of three runs, while the shaded areas represent the corresponding variances. Higher means are desirable, while smaller variances indicate better performance. The results of the sensitivity analysis experiments are shown in Fig. 15. Based on the experimental results, it can be observed that the case with an exploration variance of 0.35 leads to the worst Sim2Real transfer performance. The agent fails to explore the state/action space properly. The reward increase is slow for the case with an exploration variance of 0.05, and the achieved reward value is significantly lower compared to the cases with variances of 0.1 and 0.2. The results indicate that either too much or too little exploration can lead to poor transfer learning performance, and more importantly, the final control effect is unsatisfactory.

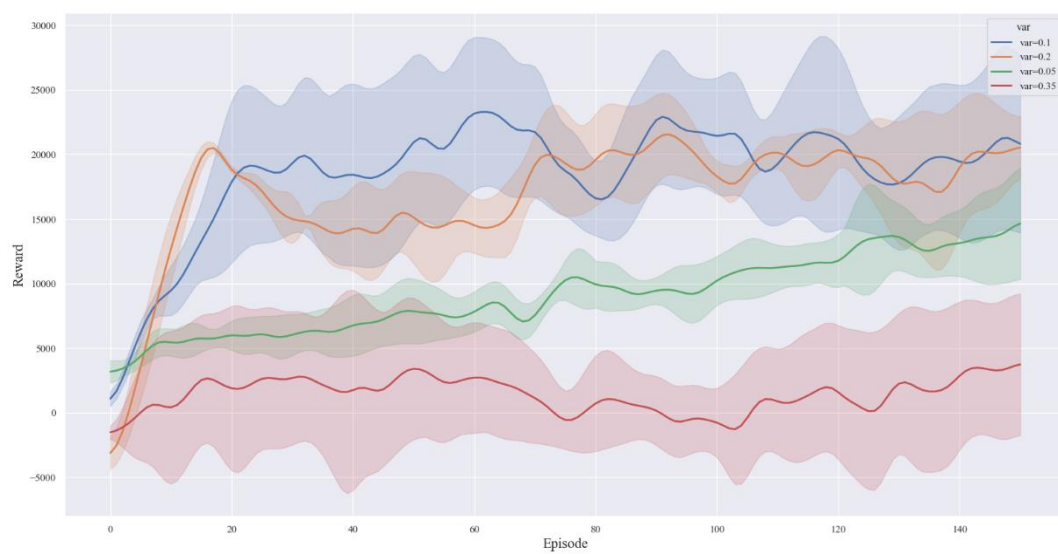

**Fig. 15 Sensitivity analysis on the role of exploration variance. The results are obtained by replacing random seeds for running 3 experiments randomly to verify the robustness of the algorithm.**

## 5. Conclusion

The ORC system has been widely employed in industrial low-grade heat recovery due to its high efficiency. This paper focuses on the superheat control of the ORC system under significant disturbances in the waste heat source and investigates the application of DRL in the Sim2Real transfer learning scenario. The main contributions of this study are summarized as follows:

1. The study utilizes the closed-loop data generated from the actual ORC system to construct a surrogate model that can quickly predict the next state based on the current state and the control signal, effectively addressing the slow system response and computational inefficiency.

2. By using a surrogate model based on LSTM, pre-training can be performed in a simulation environment without the need to interact with the actual system. This approach mitigates potential safety risks and significantly increases the pre-training speed of the DRL algorithm.

3. The surrogate model developed in this study is trained using a neural network to predict the dynamic behavior of the system. This enables the DRL agent to anticipate states that have not been encountered in the historical data, thereby improving the agent's generalization performance in multi-mode control scenarios.

4. A comprehensive exploration of the Sim2Real transfer learning scenario is conducted, emphasizing the importance of setting an appropriate level of exploration-exploitation trade-off for the DRL algorithm in terms of transfer learning performance and safety assurance.

5. The study proposes a practical and user-friendly DRL-based solution for optimization control in power cycles. This approach can be extended to other process objects in the field of energy system engineering, providing a new avenue for the application of DRL in industrial production.

## CRediT authorship contribution statement

**Runze Lin:** Conceptualization, Methodology, Software, Validation, Writing – original draft, Writing – review & editing, Visualization.

**Yangyang Luo:** Methodology, Software, Data curation, Validation, Visualization.

**Xialai Wu:** Investigation, Software, Data curation.

**Junghui Chen:** Conceptualization, Methodology, Validation, Writing – review & editing, Supervision, Funding acquisition.

**Biao Huang:** Conceptualization, Investigation, Writing – review & editing, Supervision.

**Lei Xie:** Supervision, Funding acquisition.

**Hongye Su:** Supervision, Project administration, Funding acquisition.